\documentstyle[aps,prl,floats,epsfig]{revtex}
\draft

\def\beq{\begin{equation}}
\def\eeq{\end{equation}}
\def\eea{\end{eqnarray}}
\def\bea{\begin{eqnarray}}

\def\gev{\, {\rm GeV}}
\newcommand{\gsim}{\lower.7ex\hbox{$\;\stackrel{\textstyle>}{\sim}\;$}}
\newcommand{\lsim}{\lower.7ex\hbox{$\;\stackrel{\textstyle<}{\sim}\;$}}

\def\slashchar#1{\setbox0=\hbox{$#1$}           
   \dimen0=\wd0                                 
   \setbox1=\hbox{/} \dimen1=\wd1               
   \ifdim\dimen0>\dimen1                        
      \rlap{\hbox to \dimen0{\hfil/\hfil}}      
      #1                                        
   \else                                        
      \rlap{\hbox to \dimen1{\hfil$#1$\hfil}}   
      /                                         
   \fi}                                         %

\begin{document}

\twocolumn[
\hsize\textwidth\columnwidth\hsize\csname @twocolumnfalse\endcsname

\title{Supersymmetry and the positron excess in cosmic rays}
\author{G.L.~Kane$^{a}$, Lian-Tao Wang${}^a$, James D. Wells$^{b,c}$}

\address{$^{(a)}$Physics Department, University of Michigan, 
Ann Arbor, MI 48109 \\
$^{(b)}$Physics Department, University of California, 
      Davis CA 95616\\
$^{(c)}$Lawrence Berkeley National Laboratory, Berkeley, CA 94720}
\maketitle

\begin{abstract}  

Recently the HEAT balloon experiment has confirmed an excess of
high-energy positrons in cosmic rays.  They could come from annihilation
of dark matter in the galactic halo.  We discuss expectations for the
positron signal in cosmic rays from the lightest superpartner. The
simplest interpretations are incompatible with the size and shape of the
excess if the relic LSPs evolved from thermal equilibrium.  Non-thermal
histories can describe a sufficient positron rate.  Reproducing the energy
spectrum is more challenging, but perhaps possible.  The resulting light
superpartner spectrum is compatible with collider physics, the muon
anomalous magnetic moment, Z-pole electroweak data, and other dark matter
searches.

\end{abstract}  

\pacs{PACS: 12.60 Jv, 13.40 Em , hep-ph/0108138, MCTP-01-37, LBNL-48801}

\vspace{0.18in}

]
\setcounter{footnote}{1}

{\it Recent experiments:} 
Good solutions to the cosmological dark matter problem often involve
hypothesizing a stable weakly interacting massive particle (WIMP).  
The particles
populate galactic halos providing gravitational support to the
unusual constant velocity profiles of many galaxies.

Direct experiments continue to look for WIMPs scattering
off nuclear target detectors.   Indirect experiments rely
on annihilation of ambient
WIMPs that produce
an excess above background of photons, anti-protons, positrons or neutrinos
in cosmic rays.  Each of these experiments has its unique experimental
challenges, and its unique astrophysical assumptions and uncertainties.
For example, to be successful the direct searches need 
a significant local density of WIMPs, whereas discovery of 
a monochromatic photon line from WIMP annihilations
generally requires a cusping distribution near the galactic center.
The charged particle signals ($\bar p$ and $e^+$) require 
an accurate model describing their
propagation and energy loss
from their source at WIMP annihilations in the galactic
halo to the detector on earth. 

Our imperfect understanding of the dark matter distribution and other
astrophysics uncertainties makes it impossible to predict which signal
would be the first to demonstrate evidence for WIMP dark matter.
For this reason, all the different experiments designed for this purpose
are interesting and necessary parts of a comprehensive search strategy.
Once WIMPs are found all the experiments provide information about their 
properties and help to determine the WIMP relic density.

Recently, the HEAT collaboration has found tantalizing evidence for
unexpected structure in the $e^+/(e^-+e^+)$ energy 
spectrum~\cite{Barwick:1997ig,Coutu:1999ws,Coutu:2001}. 
The first set of data from the 1994-1995 flights indicated a rise or bump
in the positron fraction at energies above about $7\gev$.
Using a different instrument, with different systematics, the HEAT
collaboration found in the data of their 2000 flight a similar
rise.  The  consistency between the data sets adds further
confidence in the measured energy distribution of the positron fraction.

{\it Attempts at a standard supersymmetry interpretation:}
One of the most compelling theories for WIMP dark matter is supersymmetry.
R-parity conserving supersymmetry naturally provides a dark matter
candidate in the lightest supersymmetric partner (LSP).  

In some models, such as ``minimal supergravity'', the LSP
is mostly bino (fermion superpartner to the hypercharge gauge boson).  
For relatively light superpartners (mass near the weak
scale), one finds in large fractions of the parameter space of 
these models that a simple thermal history calculation
will give an answer remarkably close to the $\Omega h^2\simeq 0.1$ 
needed for an acceptable cold dark matter candidate.  LSP annihilations
into positrons can then be searched for in cosmic 
rays~\cite{Tylka:1989xj,Turner:1990kg,Kamionkowski:1991ty}.
However, the standard supersymmetry model does not explain the HEAT
data, for two reasons.  

First, the positron excess is most
simply produced by LSP annihilations into $W$ 
bosons~\cite{Kamionkowski:1991ty,Feng:2001zu,Baltz:1999xv}, 
one of which
subsequently decays into a positron.  However, binos do not couple
to $W$'s and so this final state is suppressed compared
to other final states. There is still the option of producing positrons 
from cascade decays 
of the other final states of bino annihilation. For example, annihilations
into tau leptons can produce positrons from leptonic
decays of $\tau^+$ or from fragmentation of $\tau$ jets.
However, the total annihilation
rate for binos is small.
Although this is correlated with a reasonable 
$\Omega h^2$, the annihilation rate is insufficient to produce
a large flux of positrons to overcome expected backgrounds.
Therefore, the positron fraction signal is not expected to be visible, 
unless we have underestimated important astrophysical parameters 
considerably.   
The HEAT data is likely not explained
by bino LSP theories where the relic abundance of
LSPs is accurately computed from a simple thermal history of the universe.

{\it Higgsino and wino dark matter: }
What is needed to explain the HEAT signal is a large relic abundance,
a large annihilation rate, and a rising distribution of positron
fraction at energies above about 7 GeV.  Higgsino and wino LSPs may do 
this.
They couple 
at full strength to the $W$ boson and have a large annihilation rate.
As long as they have mass above $m_W$, 
higgsinos and winos
will annihilate predominantly into $WW$ final states and so can
produce a large number of high-energy
positrons from $W\to e^++X$.  This has
been discussed recently in the context of traditional supersymmetry
models with large higgsino 
fraction LSP~\cite{Feng:2001zu,Baltz:1999xv} and anomaly 
mediation with wino LSP~\cite{Moroi:2000zb}.

Since the relic abundance correlates inversely with the strength
of annihilation, there is still the worry that there will be too few of
these LSPs in the galactic halo to annihilate with each other and
produce a signal. However, that argument is based on a standard
thermal history calculation which predicts
$\Omega_{\rm LSP}\lsim 10^{-3}$.  Non-thermal sources and non-standard
cosmologies have been found to produce a significant relic abundance
independent of the thermal annihilation 
rate~\cite{Gherghetta:1999sw,Moroi:2000zb,Jeannerot:1999yn}.  
It is one of the important
conclusions of this paper that the higgsino dark matter density probably must
be understood outside the normal thermal evolution framework if the HEAT data
is indicating LSP annihilations.

{\it Details of the positron signal:} 
In order for the reader to understand our results we will 
briefly describe the assumed dark matter density profile
we use to produce expected positron fluxes from LSP annihilations.
The dark matter halo is assumed to be spherically symmetric 
isothermal sphere whose density at a position $r$ from the
galactic center is
\beq
\rho (r)=\rho_0\frac{a^2+r_0^2}{a^2+r^2}
\eeq
where $\rho_0=0.3\gev/{\rm cm}^3$ is the local LSP density,
$a=3.5$ kpc is the core radius, and
$r_0=8.5$ kpc is the distance of the earth from the galactic center.

The flux $F_{e^+}$ of positrons at the detector for higgsino or wino
can be calculated from
\beq
\frac{dF_{e^+}}{dE}=\frac{\rho_0^2}{m_\chi^2}\int d\epsilon\,
 G_{e^+}(E,\epsilon)\sum_{f} (\sigma v)_f A_{e^+}^f(\epsilon)
\eeq
where $(\sigma v)_f$ is the annihilation rate of $\chi\chi$ into
the final state $f=WW$ or $f=ZZ$, $G_{e^+}(E,\epsilon)$ is the positron
propagation Green's function,
and $A_{e^+}^f(\epsilon)$ is the
average positron energy distribution function for the final state $f$
at the source (pre-propagation).  The
$A^f_{e^+}$ functions are normalized such that
$\int d\epsilon \, A_{e^+}^{f}(\epsilon)$ is the average number of positrons
in decays of the final state $f$.  

We utilize DARKSUSY \cite{Gondolo:2000ee} for calculating the flux.
We also tested the results by simulating the 
$A_{e^+}^f(\epsilon)$ from Pythia.  The Green's function 
$G_{e^+}(E,\epsilon)$ can be extracted from 
refs.~\cite{Baltz:1999xv,Moskalenko:1999sb}.  
The numerical values we used are 
from~\cite{Baltz:1999xv,Gondolo:2000ee} 
with energy loss time $\tau_E=10^{16}\, {\rm sec}$,
and with energy-dependent diffusion constant
\beq
K(\epsilon)=6.1\times 10^{27}\, \left(\frac{\epsilon}{1\gev}\right)^{0.6}\, 
{\rm cm}^2\, {\rm sec}^{-1}.
\eeq

As indicated earlier, the thermal relic abundance is much too small
to be of cosmological significance, but non-standard mechanisms can
save the higgsino and wino as dark matter candidates. From 
here on we assume that the local density 
$\rho_0$ is made up entirely of neutralino dark matter, assuming a
non-thermal
source for the LSPs such as from late decays of very heavy 
gravitinos~\cite{Gherghetta:1999sw}.
Then we no longer need to concern ourselves with
neutralino relic abundance, since the value of $\rho_0$ 
captures all the information we need about LSP abundance in our
positron flux calculation.  An obvious consequence of this approach is
that we do not
rescale $\rho_0$ according to the thermal relic abundance calculation.

\begin{figure}[t]
\centerline{\epsfxsize=3.0truein \epsfbox{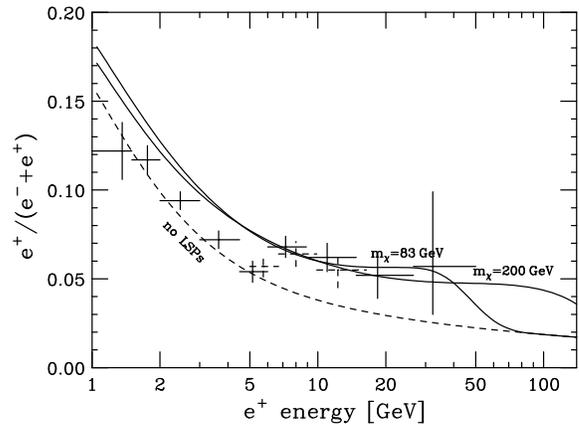}}
\vskip 0.0 cm
\caption{Positron fraction as a function of energy.  The lower dashed line
is the expected signal fraction with no LSP annihilations
for a certain set of astrophysical assumptions described in the text. The
solid lines also include the positrons and electrons from the
annihilations in the galactic halo of LSPs with mass $m_\chi=83\gev$
and $m_\chi=200\gev$, with boost factors of 2.7 and 3.9 respectively.
The 1994-1995 HEAT data is represented by the solid line cross-hairs,
and the 2000 HEAT data by the dashed line cross-hairs.}
\label{higgsino}
\end{figure}

Fig.~\ref{higgsino} shows the positron fraction energy distribution
for the HEAT data~\cite{Coutu:1999ws,Coutu:2001}, 
expected distribution with no LSP annihilations,
and expected distribution with LSP annihilations subject to the
above assumptions.  The plot is made for higgsino/gaugino mixed scenario
with LSP masses of $m_\chi =83\gev$ and $200\gev$.

We have normalized the positron distribution to $0.06$ at $10\gev$
for each value of the LSP mass.  To do this we had to arbitrarily multiply
the flux calculated with the above assumptions by a factor 
(``astrophysical boost factor'') 
of $2.7$ and $3.9$ for $m_\chi=83\gev$ and $200\gev$, respectively.
An extra boost factor less than about 10
is probably well within the astrophysical uncertainties of parameters
used to calculate the flux. This gives us confidence that low-mass LSPs
$m_W< m_\chi \lsim 200\gev$ are worth pursuing as possible interpretations 
of the
positron energy distribution and the total flux.

\begin{figure}[t]
\centerline{\epsfxsize=3.0truein \epsfbox{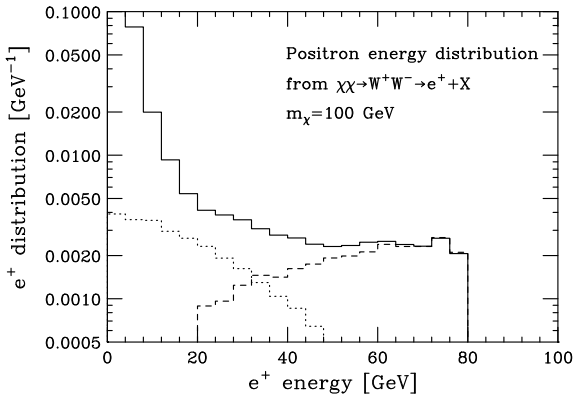}}
\vskip 0.0 cm
\caption{The solid line is
the average positron energy distribution from cascade decays
resulting from LSP annihilations into $W$-boson pairs. The mass of the 
LSP is 100 GeV in this example.  The dashed line tracks the
positrons from $W^+\to e^+$ direct decays, and
the dotted line from $W^+\to \mu^+/\tau^+\to e^+$ direct
decays.}
\label{positrons}
\end{figure}

The HEAT data  appears to show
a dip in the positron energy fraction near $E_{e^+}=7\gev$.  A dip would
indicate that a signal should have a large bump in its positron
distribution at energy above $7\gev$. $\chi\chi\to W^+W^-$ annihilations
are the best hope to produce a bump in
the positron spectrum from LSP annihilations, since $W^+\to e^+\nu$ decays
lead to a peak in the positron spectrum at high energies.  However,
there are numerous other sources for positrons in $W$ decays, including
cascades from $\tau$ and $\mu$ leptons, and decays of pions in jets.
In Fig.~\ref{positrons} we show the average positron energy distribution
from $\chi\chi\to W^+W^-$ annihilations, simulated using Pythia 
results from $e^+e^-\to W^+W^-$.  The lack of a peak in this distribution
clearly indicates that simple LSP annihilations cannot reproduce a
strong peak in the positron energy spectrum.

Therefore, in order for the LSP annihilations
to be consistent with the data we have to assume that there is no
significant dip in the data, but rather a change in slope.  
Given the error bars for the HEAT data
points, this possibility is not out of the question. 
In this case, the signal arises from an LSP-induced
positron distribution that is somewhat flatter than the background
positrons.  The prediction is best fit to the data when the number of
signal positrons starts to become a significant fraction of the total
positron rate at $E_{e^+}\gsim 5\gev$.

We briefly mention here another interpretation of the data which 
is  somewhat fine-tuned, but would be more consistent with strong 
peaking in the positron energy
spectrum.  The electron
sneutrino is stable or nearly stable if its
mass is extremely close to the LSP's.  In this case one could imagine
$\chi \tilde \nu^* \to W^-e^+$ annihilations with the positron energy
peaked at $E_{e^+}=m_\chi (1-m^2_W/4m_\chi^2)$.  
Numerically, to get a sharp peak at about 8 GeV requires
$m_{\tilde{\nu}} + m_{\chi} \approx m_W+10\gev$. Whether the neutralino
or sneutrino is the lightest would not be important.
We have not carefully studied this possibility, although we recognize that
the LEPII collider data would severely constrain it, and maybe even rule 
it out.

{\it Correlated phenomena:} 
In the previous sections we have concluded that 
\begin{itemize}
\item  Traditional supersymmetry with thermal relic abundance near
     $\Omega h^2\simeq 0.1$ cannot yield an excess of
       positrons above background because the annihilation rate is too low, 
\item A higgsino or wino LSP with mass $m_W<m_\chi \lsim 200\gev$
could yield a significant excess of positrons
 above background provided
 the relic abundance is from a non-thermal source,
\item A generic
higgsino or wino interpretation is consistent with the HEAT data 
only if no strong dip is present,
\item Uncertainties in the data and in 
astrophysical processes such as positron
production, propagation and modulation means we may 
not need a new-physics interpretation of
the HEAT data, although our current understanding suggests we do.
\end{itemize}

If the higgsino or wino interpretation of HEAT data is correct, we should
expect other correlating phenomena that can be measured and quantified.
First, it is well-known by now that the recent excess~\cite{Brown:2001mg} 
in the muon anomalous magnetic moment is consistent with light 
supersymmetry~\cite{Everett:2001tq,Feng:2001tr,Martin:2001st}.  
Particularly, a light
higgsino and large $\tan\beta$ are
helpful to get a large supersymmetric correction, since the
higgsino-smuon-muon vertex is a $\tan\beta$ enhanced chirality flip.
The supersymmetric contribution to $\delta a_\mu$ with higgsino or
wino LSP is
\beq
\delta a_\mu \lsim 14 \tan \beta \left( 
\frac{100\gev}{m_{\chi}}\right)^2\times 10^{-10}.
\label{asusy}
\eeq

Equality in eq.\ (\ref{asusy}) is attained when all other sparticle masses
are very close to the LSP mass.
The measurement minus the Standard Model contribution is
$\delta a_\mu = (41\pm 16)\times 10^{-10}$.  Therefore, moderate $\tan\beta$ 
and low mass superpartners have no difficulty recovering
the central value for the measured $\delta a_\mu$.

Colliders can also search for higgsino and wino cold dark matter.  These
searches are notoriously 
difficult because there is no guarantee that
visible superpartners have mass close to the LSP mass and are therefore
accessible by the colliders.  In the case of higgsinos and winos, there
are charged particles nearby, $\tilde H^+$ or $\tilde W^+$; however,
they are almost degenerate in mass to the LSP.  The production
of $e^+e^-\to \tilde H^+\tilde H^-$ may be high, but the final state
of two soft pions from $\tilde H^\pm\to \tilde H^0 \pi^\pm$
is very difficult to find. Searches have been conducted, and the mass
limits for these sparticles are about $m_W$~\cite{Abreu:1999qr}, 
just below the interesting
region for the HEAT signal.  Future lepton colliders will have a much
higher mass reach, and hadron colliders will be useful if other superpartners
are produced~\cite{Gunion:2001fu}.

Other astrophysics experiments may also see evidence for
wino or higgsino cold dark matter. Nuclear target detectors have limits
that are already sensitive
to wino or higgsino LSPs in some parts of parameter 
space~\cite{Murakami:2000me}.  However,
these limits depend on squark masses, heavy Higgs boson masses, etc.\
which feed into the spin-independent nucleon-LSP scattering cross-section,
and which have little to do with the positron fraction prediction.
Therefore, it is difficult to predict how sensitive next generation
cryogenic detectors will be to light higgsinos and winos. A possible
signal not in conflict with our results
has been reported at DAMA~\cite{Bernabei:2000qi,Belli:2000nz}.

On the other hand, loop-induced annihilations of
$\chi\chi\to \gamma\gamma$ are very high for higgsinos and winos.
One therefore expects a monochromatic photon signal to arise from
annihilations of winos and higgsinos in the galactic 
halo~\cite{mrenna-wells,Ullio:2001qk}.  
As mentioned at the beginning,
the astrophysical uncertainties of this calculation are quite different
than the positron fraction calculation.  Therefore, it is difficult
to predict if experiments such as GLAST will see a signal, but we
do expect so if the HEAT results are due to LSP annihilations.

Finally, it has been suggested recently that the precision electroweak
data is more consistent with light
superpartners~\cite{Altarelli:2001wx}.  Sneutrino masses below $m_W$ 
and slepton masses just above the experimentally allowed region are the
most important requirements for the successful fit to data.  Light
gauginos are also helpful, but not as critical.  Therefore our
interpretation
of the HEAT data, which requires superpartner masses near $m_W$,
is not only compatible with the precision electroweak data, but may
be encouraged by it. This is another reason why the LSP interpretation is 
worthwhile pursuing even though it has difficulty reproducing
the precise structure of the data.

\vspace{0.2in}
\noindent
{\it Acknowledgments: }
This work was supported in part by National Science Foundation, the
Department of Energy, and the Alfred P. Sloan Foundation. We appreciate 
stimulating discussions with G. Tarle, S.~Mrenna, and D.~Chung, and help from 
T. Wang. 

\def\Journal#1#2#3#4{{#1} {\bf #2}, #3 (#4)}
\def\add#1#2#3{{\bf #1}, #2 (#3)}

\def\NPB{{\em Nucl. Phys.} B}
\def\PLB{{\em Phys. Lett.}  B}
\def\PRL{{\em Phys. Rev. Lett.}}
\def\PRD{{\em Phys. Rev.} D}
\def\PR{{\em Phys. Rev.}}
\def\ZPC{{\em Z. Phys.} C}
\def\SJNP{{\em Sov. J. Nucl. Phys.}}
\def\AnnP{{\em Ann. Phys.}}
\def\JETPL{{\em JETP Lett.}}
\def\LMP{{\em Lett. Math. Phys.}}
\def\CMP{{\em Comm. Math. Phys.}}
\def\PTP{{\em Prog. Theor. Phys.}}

\end{document}